\documentclass[aps,pre,twocolumn,floatfix,nofootinbib,showpacs,amsmath,amssymb]{revtex4}

\usepackage{times}
\usepackage[sort&compress]{natbib}
\usepackage[dvips]{graphicx}
\usepackage[dvips]{color}
\begin{document}

\title{Dynamics of the Chemical Master Equation, a strip of  chains of
equations in d-dimensional space. }

\author{Vahe Galstyan$^{1}$}
\author{David B. Saakian$^{2,3,4}$}
\email{saakian@yerphi.am}
 \affiliation{$^1$Quantum College,
Bagratunyats 23/2, Yerevan 0046, Armenia } \affiliation{$^2$Yerevan
Physics Institute, Alikhanian Brothers St. 2, Yerevan 375036,
Armenia} \affiliation{$^3$Institute of Physics, Academia Sinica,
Nankang, Taipei 11529, Taiwan}
 \affiliation{$^4$National Center for
Theoretical Sciences (North): Physics Division, National Taiwan
University, Taipei 10617, Taiwan}

\begin{abstract}
 We investigate the multi-chain version of the Chemical Master
Equation, when there are transitions between different states inside
the long chains, as well as transitions between (a few) different
chains. In the discrete version, such a model can describe the
connected diffusion processes with jumps between different types. We
apply the Hamilton-Jacobi equation to solve some aspects of the
model. We derive exact  {(in the limit of infinite number of
particles)} results for the dynamic of the maximum of the
distribution and the variance of distribution.
\end{abstract}
 \maketitle

\section{Introduction}

The Chemical Master Equation (CME)
\cite{ga},\cite{ka92},\cite{qi08},\cite{qi10} is the main tool to
investigate chemical reactions with few molecules. This statistical
physics model describes also  the branching-annihilation processes
and (in strip version, considered in the current article) the
discrete version of the connected diffusion.

For reactions with few molecules the number of molecules fluctuate,
and we can describe the reaction process through the probabilities
of having a given number of molecules.
 While
working with D types of molecules, we work with the probabilities of
having given integer number of molecules $P(X_1...X_D,t)$, thus the
variables are located at the nodes of a D-dimensional lattice, where
integer $X_i$ describe the number of molecules of the i-th type. One
 solves a linear system of differential equations for $P(X_1...X_D,t)$
 to define the probability of having the given numbers of molecules at time t.
 The CME is a linear equation, like the Shroedinger equation for the
 wave function. We assume that the system does not interact with the environment,
 and during the measurement we observe just some collection of
 integers for the different types of molecules, defined by probabilities
 $P(X_1...X_D)$.

 While the numerical solution of CME is possible using the Gilepsee
algorithm, the analytical investigation of CME is very important.
The problem was solved exactly in the infinite system size (the
maximal number of particles) limit for $D=1$ using quantum mechanics
or mapping to the Hamilton-Jacobi equation (HJE)
\cite{hu88},\cite{dy94},\cite{ka04},\cite{es09},\cite{as10}. In
\cite{sa11} an implicit expression was obtained for the dynamics of
the population distribution variance, following the HJE method of
\cite{sa07},\cite{ka07},\cite{sa08a}. To obtain an exact complete
solution of CME in dimensions $D\ge 2$ is a very hard problem and
only a few solutions are known, see \cite{kr11}.

Sometimes the CME is organized as a strip of systems of equations
\cite{qi10}-\cite{ra11}. In \cite{qi10} strips of one dimensional
chains of equations (the variables are located on 1-D axes and the
time derivative of some variable depends on its neighbors) are
considered as a model of coupled diffusion processes: a situation in
2 dimensional space, when there is a diffusion in one dimension, and
jump process in the other dimension (continuous diffusion later is
replaced by discrete jumps). \cite{ev11} considers a strip of three
chains of equation, each of chains defined in 2-dimensional space,
identifying the model with genetic switch phenomenon. Again we have
a  diffusion, completed by jump process.

In the current article we formulate the following general problem:
we investigate the CME, organized as a strip of d chains of
equations, each defined in D-dimensional space with a maximal number
of a given type of molecule N, where $d\ll N$.

We will also give differential equations, defining the solution of
the moments of distributions of the model for the general $d,D$
case. Two mathematical methods have been applied to solve the CME in
the limit of large number of molecules: diffusion approximation and
the HJE. The HJE method allows to apply the methods of Hamilton
mechanics \cite{la}, the Hamilton equation for the characteristics
\cite{ev02},\cite{me98}. For the recent introduction to the method
see \cite{qi11}. In this article we consider the situation when the
HJE method is valid while describing the solution of the CME.

Let us first review known results of the CME. In the simplest case
($D=1,d=1$) and one step processes ($K=1$), we have a system
 of equations, see Fig. 1 for illustration,
 \begin{figure}
\centerline{\includegraphics[width=0.95\columnwidth]{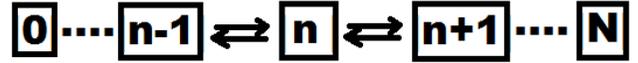}}
\caption{The CME for the case $d=1,D=1,K=1$. Possible states of the
system (indicated as boxes with the number of molecules inside) are
located on the 1-d space ($D=1$). There are transitions only between
neighbors ($K=1$). There is only one chain ($d=1$). The back
transition rates is  $R_{-1}$ and the forward transition rate is
$R_{1}$, see Eq. (1). } \label{fig2}
\end{figure}

\begin{eqnarray}
\label{e1}
\frac{dP(X,t)}{Ndt}=R_1(X-1)P(X-1,t)+\nonumber\\
R_{-1}(X+1) P(X+1,t)+R_0(X)P(X,t)\nonumber\\
R_0(X)=-(R_1(X)+R_{-1}(X)),
\end{eqnarray}
where $0\le X\le N$, and N is the maximal possible number of
molecules, a large number. $R_1(X)$ is the forward 1-step transition
rate, and $R_{-1}(X)$ is the back 1-step transition rate.

 A similar
equations has been formulated for the
 virus evolution problems.
In case of virus evolution models \cite{hi96},\cite{ba01}, there is
inhomogeneous term (fitness function) on the right hand side of
equation, and $P(X)$ describe the density of different types of
viruses.

 While in virus evolution models  one can consider a mixed state at the
 start, in case of the CME, by Eq.(1), there is a definite value n for  X at the start.
 Therefore
\begin{eqnarray}
\label{e2} P(X,0)=\delta_{X,n}
\end{eqnarray}
If during an elementary reaction event there is a consumption or
birth of several molecules with maximal consumption (birth) number
K, then our system of equations is modified:
\begin{eqnarray}
\label{e3} \frac{dP(X,t)}{Ndt}=\sum_{-K\le m\le K} R_m(X-m) P(X-m,t)
\end{eqnarray}
Here $R_{\pm m}(X), m\ne 0$ are the rates of m-step transitions,
therefore are nonnegative numbers, while $R_0 (X)= -\sum_{-K \le m\le K}R_m(X)$ (a generalization of
the expression by Eq.(1)) ensures a probability conservation
condition, see Fig. 2 for an illustration. The probability rate in
the CME are defined via
 reaction rates in case of infinite number of molecules, see \cite{qi08}.
 Different rates $R$-s in our equations describe different
 elementary reaction events.
\begin{figure}
\centerline{\includegraphics[width=0.95\columnwidth]{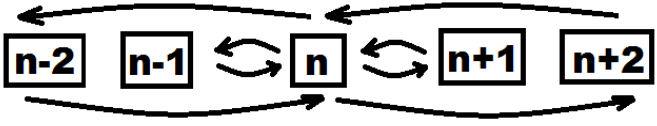}}
\caption{ The CME for the case $d=1,D=1,K=2$. Possible state of the
system are located on the line ($D=1$). We illustrate possible
transitions from the position $n$. There are 2-step transitions
($K=2$). There is only one chain ($d=1$). The back 1-step transition
rate is $R_{-1}$ and the forward 1-step transition rate is $R_{1}$.
The 2-step back transition rate is $R_{-2}$ and the
 two-step forward transition rate is $R_{2}$} \label{fig3}
\end{figure}
 In the literature, models with $K\le 3$ has been considered.
 The derivative $dP(X)/dt$ depends on $P(X')$ for the neighbors and
 that\rq s why we name our system of equations as a "chain" of
 equations.

Consider the case of D-dimensional space,
\begin{eqnarray}
\label{e4} \frac{d{P(\vec X,t)}}{Ndt}= \sum_{l=1}^{D} \sum_{n_l=-K}^{K}R_{\vec
n}(\vec X-\vec n)P(\vec X-\vec n,t)
\end{eqnarray}
Here $\vec X$ is a D-dimensional vector, where $D$ is the number of
different type of molecules participating in the reaction,
and for $\vec n\ne 0$ the nonnegative number $R_{\vec n}$ describe
the probabilities of the jump with the change of the number of the
l-th type via $n_l$, $1\le l\le D$, and negative number $R_0$
ensures the probability conservation. As an illustration consider
the case $D=4$. The process, when the molecules of types 1, 2, 3
collide giving a birth of 3 molecules of type 4 is described via a
coefficient
 $R_{-1,-1,-1,3}(x_1+1,x_2+1,x_3+1,x_4-3)$.
If these three molecules just catalyze the birth of one molecule of
type 4 the process is described via coefficient
$R_{0,0,0,1}(x_1,x_2,x_3,x_4-1) $.

If there are n reactants, then the coefficient in the CME $ R$ is
connected with the coefficient in the chemical kinetic equation
$\hat R$ with the scaling $R=\hat R/N^n$, see \cite{qi10}.

The Hamilton-Jacobi equation (HJE) is a powerful method to
investigate the CME at large values of N. For illustration consider
Eq.(1).  For the case $N\gg 1$ we assume an ansatz
\begin{eqnarray}
\label{e5} P(X,t)=\exp[Nu(x,t)],\nonumber\\
x=\frac{X}{N}.
\end{eqnarray}
The ansatz by Eq.(5) is similar to the WKB approximation in quantum
mechanics \cite{as10}. Eq.(5) gives  with $1/N$ accuracy:
\begin{eqnarray}
\label{e6} P(X\pm 1,t)=\exp[Nu(x,t)\pm u'(x)],\nonumber\\
x=\frac{X}{N},
\end{eqnarray}
and one obtains \cite{dy94}:

\begin{eqnarray}
\label{e7} \frac{\partial u(x,t)}{\partial t}+H(x,u') = 0\nonumber\\
-H(x,p)=r_-(x)e^{-p}+r_+(x)e^{p}-(r_+(x)+r_-(x)) \nonumber\\
p = \frac{\partial{u}}{\partial{x}}
\end{eqnarray}
where we denote $r_-(x)=R_{-1}(Nx),r_+(x)=R_{+}(Nx)$. Equation (7)
is a partial differential equation of a special form, the
Hamilton-Jacobi equation. According to the standard handbook of
classical mechanics \cite{la}, it can be solved using the solution
of Hamilton's equation for $x(t),p(t)$, where $x(t)$ describes the
characteristic curves of HJE \cite{me98},\cite{ev02},\cite{sa08a}. {
The N in the dominator of the left hand side of Eqs.(3),(4) is
written arbitrary. We are choosing a proper scaling $N^n$ with some
integer n, to get a smooth, N independent functions r(x) at the
limit $N\to \infty$. We can consider the more general case $N^n$,
$n>1$, then rescaling $t\to t/N^{n-1}$ return to the case of
article.}The HJE approach allows to define the whole distribution
($P(X)$ or $u(x)$). In the following sections we will derive the HJE
version of the CME for more involved situations than Eq.(3). In this
work we are mainly interested in the dynamics of the maximum of
distribution (the point x where $u_x'(x,t)=0$) and in the variance
of the distribution $P(X,t)$, given by $u''(x,t)$ at the point of
maximum. For our purposes there is no need to apply a powerful
method of characteristics. We will use directly the HJE. We are
interested in the solution $u(x,t)$ near the maximum of the
distribution, $x=y(t)$ in our notation. We can derive exact
relations by expanding in the HJE the solution in degrees of
$(x-y(t))$. Such a method has been applied in \cite{sa11}, where we
derived the Van Kampen's large system size expansion method from
HJE.

In Section II-A we derive HJE for the $d>1,D=1$ case. While we
derive the HJE to describe the full distribution, it is more
informative to calculate the motion of the maximum and the variance.
We derive corresponding ODEs in Sections II-B and II-C. In Section
II-D we considered an example of $D=1,d=2$. In Section III-A we
consider the strip of D-dimensional equations. In the Appendix we
present the equations to define the dynamics of the third moment of
distribution, investigation of evolution insertion-deletion model
\cite{sa08}. In section III-B we give the finite size corrections to
the bulk solution, following to \cite{sa07},
\cite{es09},\cite{as10}.

\section{The strip of one dimensional chains}
\subsection{Derivation of HJE}

Here we consider the CME, when the probability depends on two
integer coordinates, $l$ and $X$, thus we work with $P_l(X,t)$. There
is a  probability conservation condition
\begin{eqnarray}
\label{e8} \sum_{1\le l\le d}\sum_{0\le X\le N}P_l(X,t)=1
\end{eqnarray}
In Eq.(8) the maximal number of molecules is constrained by N, a
reasonable constraint in case of chemical reaction in a finite
volume, and $d\ll N$.

In principle only one boundary condition at $X=0$ can be present or
some large parameter $N$. In both cases HJE method can be applied.

 We consider a large N limit:
\begin{eqnarray}
\label{e9} N\to \infty.
\end{eqnarray}
There is a system of equations
\begin{eqnarray}
\label{e10} \frac{dP_l(X,t)}{Ndt}=\sum_{1\le n\le d}\sum_{-K\le m\le
K}R_{lnm}(X-m) P_n(X-m,t).
\end{eqnarray}
There are transitions along the
states of one chain (maximal jump is K-step), described by
nonnegative coefficients $R_{llm}, m\ne 0$ and there are jumps
between different chains, described by nonnegative $R_{lnm}$ with
$l\ne n$. There are negative $R_{ll0}$,  and there is an equation
\begin{eqnarray}
\label{e11}\sum_{1\le l\le d}\sum_{-K\le m\le K}R_{lnm}(X)=0
\end{eqnarray}
to support the probability conservation condition.

The system is slightly modified near the borders $X=0$ and $X=N$. In
principle, we can use a one-sided boundary condition, but for our
case of solution, described by the HJE, the boundary conditions are
irrelevant.

Let as assume that at the limit $N\to \infty$ (condition a.), the
rates are described via smooth functions
\begin{eqnarray}
\label{e12}R_{lnm}(X)=r_{lnm}(x),\nonumber\\
x=\frac{X}{N}
\end{eqnarray}
We consider the models under the condition
\begin{eqnarray}
\label{e13}det\{A_{ln}\}=0,\nonumber\\
A_{ln}=\sum_{-K\le m\le K}r_{lnm}(x)
\end{eqnarray}
Eq. (13) is a
consequence of probability conservation condition during the
transitions between different chains at the position X, see model by
Eq.(31) as an example.

 Let us consider the following
ansatz:
\begin{eqnarray}
\label{e14} P_l(X)=v_l\exp[Nu(x,t)],\nonumber\\
x=\frac{X}{N}.
\end{eqnarray}
{\bf We assume that the maximum of different components $P_l(X)$ are
near the same $X$}. Thus, if the maximum is at the points $X_1$ and
$X_2$, then
\begin{eqnarray}
\label{e15}  \frac{|X_1-X_2|}{N}\ll 1.
\end{eqnarray}
The smoothness of functions $r_{lnm}$ in Eq.(12) is crucial for the
ansatz Eq.(14). In \cite{sa08a} we considered an example, when the
exponential ansatz $p \sim \exp[Nu(x)]$  is invalid for non-smooth
functions of transition rates.

We put also the condition b:
\begin{eqnarray}
\label{e16}  d\ll N.
\end{eqnarray}
Then we have the following system of equation:
\begin{eqnarray}
\label{e17} v_l\frac{\partial u}{\partial t}=\sum_{1\le n\le
d}\sum_{-K\le m\le K}r_{lnm} (x)v_ne^{-mu'}.
\end{eqnarray}
We dropped the terms $dv_l/dt$, assuming $d\ll N$, where $N$ is the
system size or some large parameter.

As the system of equations for $v_l$ should be consistent, we have
the condition:
\begin{eqnarray}
\label{e18} det[M_{ln}(x,u')-q\delta_{ln}(x)]=0,\nonumber\\
M_{ln}(x,u')=\sum_{-K\le m\le K}r_{lnm}(x) e^{-mu'},\nonumber\\
q=\frac{\partial u}{\partial t}.
\end{eqnarray}
We assumed that {\bf $M_{ln}(x,u')$ is a (smooth) analytical
function of $x$, and all the terms of $M$ have the same, zero degree
of N (the case of the same, non-zero degree of N could be
investigated after re-scaling of time)}. In this case we can
consider $N\to \infty$ limit in a proper way. We denote $x=X/N$.

Eq.(13) gives
\begin{eqnarray}
\label{e19} det[M_{ln}(x,0)]=0
\end{eqnarray}
Eq. (18) is our main equation. It has a high degree of $q=
\frac{\partial u}{\partial t}$ and, therefore, it defines a standard
HJE with the first degree of q as in Eq.(7) and corresponding
Hamiltonian implicitly. To construct the full solution $u(x,t)$ we
have to find such an explicit Hamiltonian. Fortunately, there is no
need for an explicit Hamiltonian to investigate the properties of
the solution near the maximum of distribution.
 Expanding the
left hand side of Eq.(18) in the degrees of q, we re-write Eq.(18)
as
\begin{eqnarray}
\label{e20}\sum_{l=0}^d(-q)^lH_l=0
\end{eqnarray}
Eq.(19) gives:
\begin{eqnarray}
\label{e21} H_0(x,p)|_{p=0}=0 \nonumber\\
p \equiv \frac {\partial u}{\partial x}
\end{eqnarray}
 In
investigating the CME we are first interested in the maximum of the
distribution, then in the variance.
\subsection{The dynamics of the maximum of distribution}
Let us first consider the dynamics of the maximum. While calculating
the dynamics of the maximum, it is enough to consider 0-th and first
terms in Eq.(20).

We assume the ansatz
\begin{eqnarray}
\label{e22} u=-\frac{V(t)}{2}(x-y(t))^2.
\end{eqnarray}
Let us differentiate Eq.(20) with respect to x, and consider the
point $x=y(t)$. The higher terms of $q^l$ disappear, as $q\sim
(x-y(t))$. We obtain:
\begin{eqnarray}
\label{e23} -V\frac{\partial H_0(x,p)}{\partial p}|_{p=0}-q'_xH_1 =
0.
\end{eqnarray}
 Using $H_0=det[M_{ln}]$,
$H_1=-\frac{d}{dq}det[M_{ln}(x,u')-q\delta_{ln}]|_{q=0}$, we obtain
\begin{eqnarray}
\label{e24} \frac{dy(t)}{dt}=-\frac{ H_{0,p}'(y,0)}{H_1(y,0)}\equiv
b(y).
\end{eqnarray}
Eq.(24) defines the dynamics of concentration in case of the
infinite number  of molecules. It is the first step in the
investigation of the CME.

\subsection{The dynamics of the variance}
To investigate the dynamics of the variance, we should consider the
$q^2$ term in Eq.(20). We will consider the solution near $x=y(t)$.
Dropping the  higher terms in $(x-y(t))$, we obtain:
\begin{eqnarray}
\label{e25} \frac{d^2}{dx^2}[q^2H_2-qH_{1}+H_{0})|_{p=0}=0.
\end{eqnarray}
Using $\frac{dH(x,p)}{dx}=H'_x+H'_p p'_x$,  we  calculate
\begin{eqnarray}
\label{e26}  \frac{d^2H(x,p)}{d^2x}=H''_{xx}+2H''_{xp}p'_x+H'_p
p''_{xx}+H''_{pp}(p'_x)^2.
\end{eqnarray}
Putting Eq.(26) into Eq.(25), we obtain
\begin{eqnarray}
\label{e27}
\frac{d^2}{dx^2}[q^2H_2-q H_1 + H_0]= \nonumber\\
2(q'_x)^2 H_2-q''_{xx} H_1 - 2q'_x [H'_{1x}+H'_{1p}p'_x]\nonumber\\
+H''_{0xx}+2H''_{0xp}p'_x+H'_{0p} p''_{xx}+H''_{0pp}(p'_x)^2=0.
\end{eqnarray}
Using the ansatz (22), we derive from Eq.(27)
\begin{eqnarray}
\label{e28}
\dot{V}H_1+V^2[2b^2H_2+2bH'_{1p}+H''_{0pp}]+\nonumber\\
+V[-2bH'_{1x}-2H''_{0xp}]+H''_{0xx}=0.
\end{eqnarray}
 We use the notation $b(y)$ from
Eq.(18). Using  that $H''_{0xx}=0,-b H_1-H'_{0p}=0$, we obtain
\begin{eqnarray}
\label{e29}
\dot{V}H_1+V^2[2b^2H_2+2bH'_{1p}+H''_{0pp}]+\nonumber\\
+V[-2bH'_{1x}-2H''_{0xp}]=0.
\end{eqnarray}
Using $dy/dt=b(y)$, and denoting $Q=1/V$, we finally have
\begin{eqnarray}
\label{e30} \frac{dQ}{dy}=A(y) +QB(y)\nonumber\\
B(y)=\frac{[-2bH'_{1x}-2H''_{0xp}]}{b(y)H_1(y,0)},\nonumber\\
A(y)=\frac{[2b^2H_2+2bH'_{1p}+H''_{0pp}]}{b(y)H_1(y,0)}.
\end{eqnarray}
Eq.(30) defines the dynamic of the variance
$<(x-y(t))^2>=\frac{Q}{N}$.

\subsection{The case of two 1-D chains, the genetic switch model} Consider the case
\cite{ra11}:
\begin{eqnarray}
\label{e31} \frac{dP_{1,n}}{dt}=kN[P_{1,n-1}-P_{1,n}]+\rho [(n+1)P_{1,n+1}-nP_{1,n}]\nonumber\\
-nhP_{1,n}+NfP_{2,n}\nonumber\\
\frac{dP_{2,n}}{dt}=kN[P_{2,n-1}-P_{2,n}]+\rho [(n+1)P_{2,n+1}-nP_{2,n}]\nonumber\\
+nhP_{1,n}-NfP_{2,n}
\end{eqnarray}
Thus, there are back ($\rho n$) and forward ($KN$) transition rates
in the 1-d chains, as well as $nh$ and $Nf$ transition rates between
different chains. Fig. 3 illustrates our CME.
\begin{figure}
\centerline{\includegraphics[width=0.95\columnwidth]{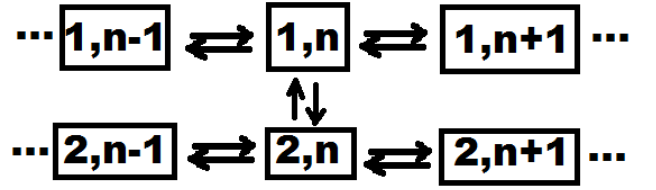}}
\caption{The CME given by Eq.(31) for the case $d=2,D=1,K=1$.
Possible state of the system are located on the two 1-d lines
($D=1$). There are only one-step transitions inside the chains
($K=1$). There are two chains ($d=2$). The backward transition rate
is $\rho n$ and the forward transition rate is $KN$. The transition
rate from the 1-st chain to the second is $nh$ and from the second
chain to the first is $Nf$} \label{fig3}
\end{figure}

The system is modified at the boundaries:
\begin{eqnarray}
\label{e32} \frac{dP_{1,0}}{dt}=-kNP_{1,0}+\rho
P_{1,1}+NfP_{2,0}\nonumber\\
\frac{dP_{2,0}}{dt}=- k NP_{2,0} +\rho P_{2,1}-fNP_{2,0}
\nonumber\\
\frac{dP_{1,N}}{dt}=kN P_{1,N-1}-\rho NP_{1,N}
-NhP_{1,N}+fNP_{2,N}\nonumber\\
 \frac{dP_{2,N}}{dt}= k N P_{2,N-1}
-\rho
NP_{2,N}+NhP_{1,N}-fNP_{2,N}\nonumber\\
\end{eqnarray}
Eq.(18) gives
\begin{eqnarray}
\label{e33}M_{11}=k(e^{-u'}-1)+\rho x(e^{u'}-1)
-xh\nonumber\\
M_{12}=f\nonumber\\
M_{22}=k (e^{-u'}-1)+\rho x(e^{u'}-1)
-f\nonumber\\
M_{21}=x h.
\end{eqnarray}

Thus we define the function  $b(y)$ for the dynamic of the maximum:
\begin{eqnarray}
\label{e34} \frac{dy}{dt}=b(y)\equiv k-\rho y
\end{eqnarray}
and functions $A,B$ to define the dynamic of the variance:
\begin{eqnarray}
\label{e35} A(y)= \frac{k+\rho y}{k-\rho y} \nonumber\\
B(y)= -\frac{2\rho}{k-\rho y}.
\end{eqnarray}
Our analytical results are in excellent agreement with numerical
solutions of the original system (31) (see Fig. (4,5)).

 {Let us derive the time dependence of the ratio $\frac{v2}{v1}$ at the maximum of distribution. From Eq.(18) we have for our case:}
\begin{eqnarray}
\label{e36} v_1 M_{11}|_{x=y(t)} + v_2 M_{12}|_{x=y(t)} = 0
\end{eqnarray}
 {Using Eq. (33) we get:}
\begin{eqnarray}
\label{e37}
\frac{v_2}{v_1}|_{x=y(t)} = \frac{y(t)h}{f}
\end{eqnarray}
The comparison of the numerics with our analytical results is given
in Fig. 6. While $y(t)$ and $Q(t)$ follow to theoretical results
immediately, after period of time $T\ll 1$, the $v_2/v_1$ follows to
our theoretical result after period of time $T\ll N$. We derived
$y(t),Q(t)$ using the expressions of $P_1(X)+P_2(x)$, thats why
$y(t),Q(t)$ follow the theoretical dynamics rather faster than
$v_2/v_1$. As $y(t),Q(t)$ do not depend on the ratio $v_2(0)/v_1(0)$
of initial values of probabilities at the  maximum point, we don't
give these values in the Figures 4,5.

\begin{figure}
\label{fig4}
 \large \unitlength=0.1in
\begin{picture}(42,22)
\put(0, -5.0){\includegraphics{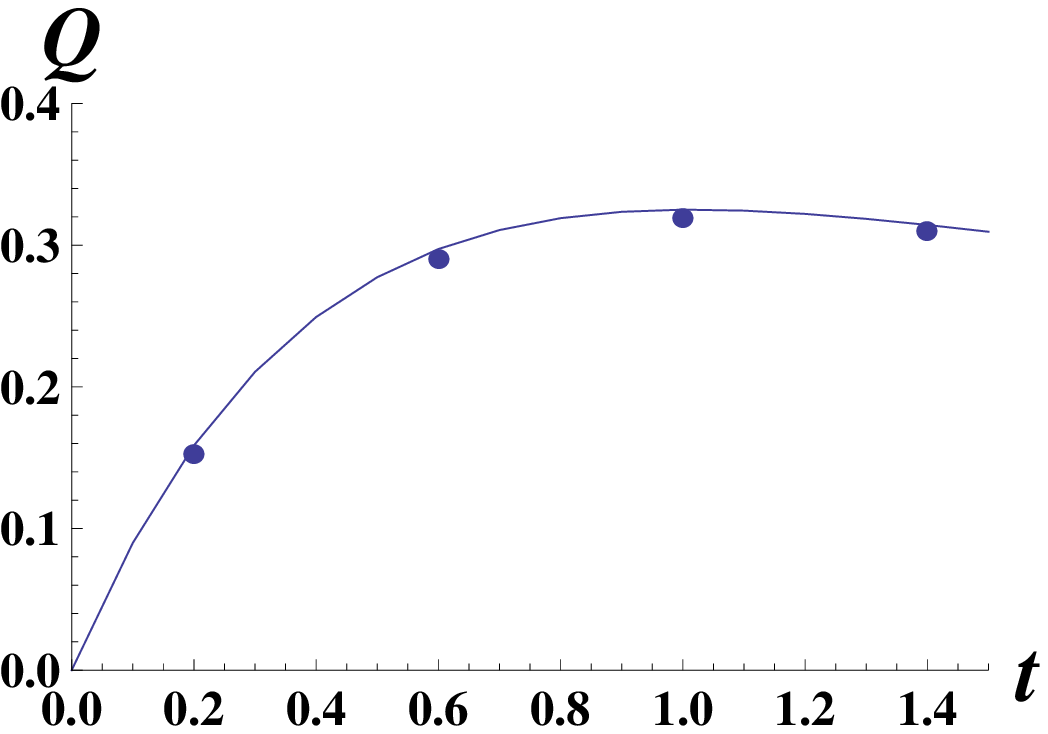}}
\put(0,10.7){\includegraphics{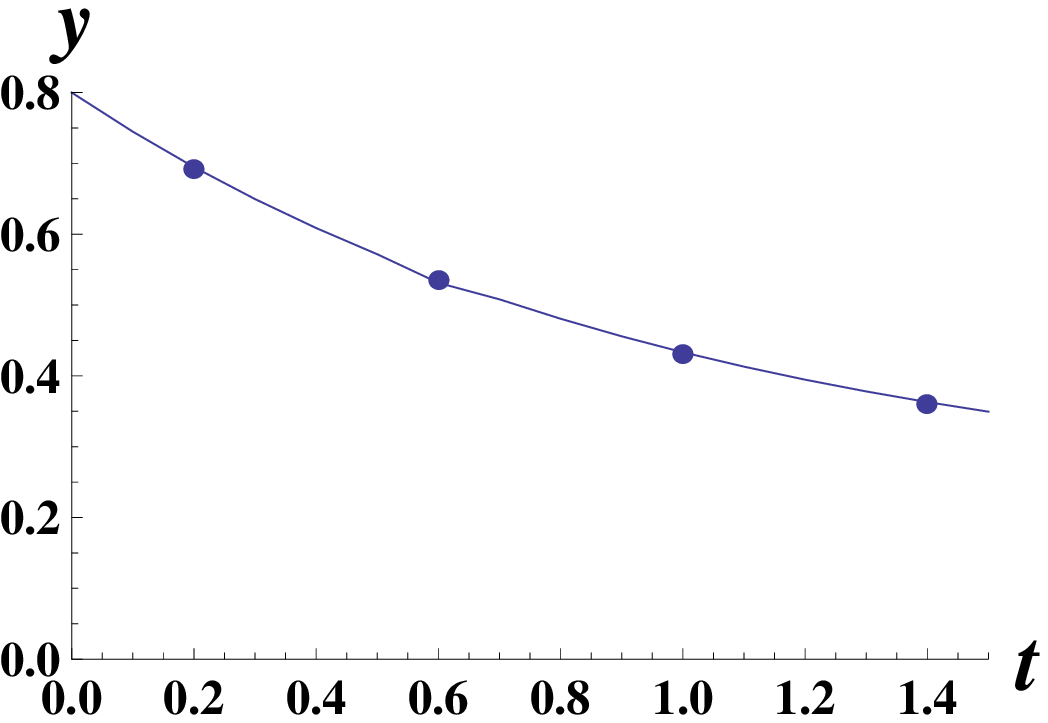}}
\put(2.7,1.0){\small{b.}} \put(2.7,20.5){\small{a.}}
\end{picture}
\vskip 10 mm \caption{The model by Eq.(31), $N=100, \rho =1, h
=0.005, f =0.02$. The direct solution of the system of
equations (smooth line) versus the analytical results (solid
circles). $y(t)$ describes the point of maximum,$1/Q(t)$ is the
variance of distribution.}
\end{figure}

\begin{figure}
\label{fig5}
 \large \unitlength=0.1in
\begin{picture}(42,22)
\put(0, -5.0){\includegraphics{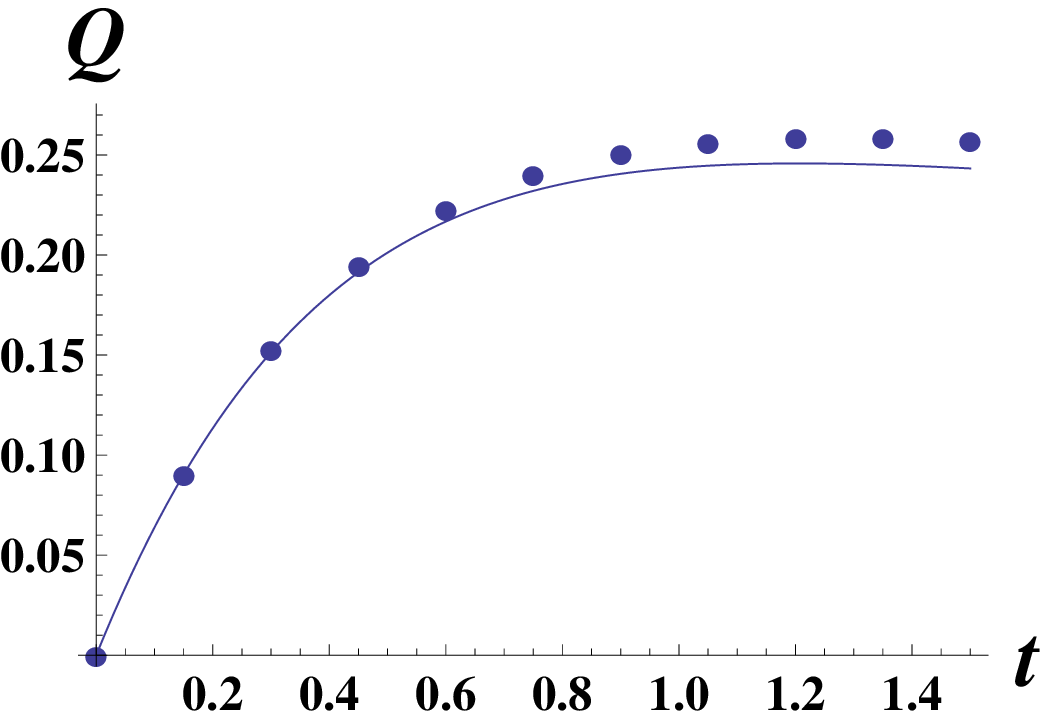}}
\put(0,10.7){\includegraphics{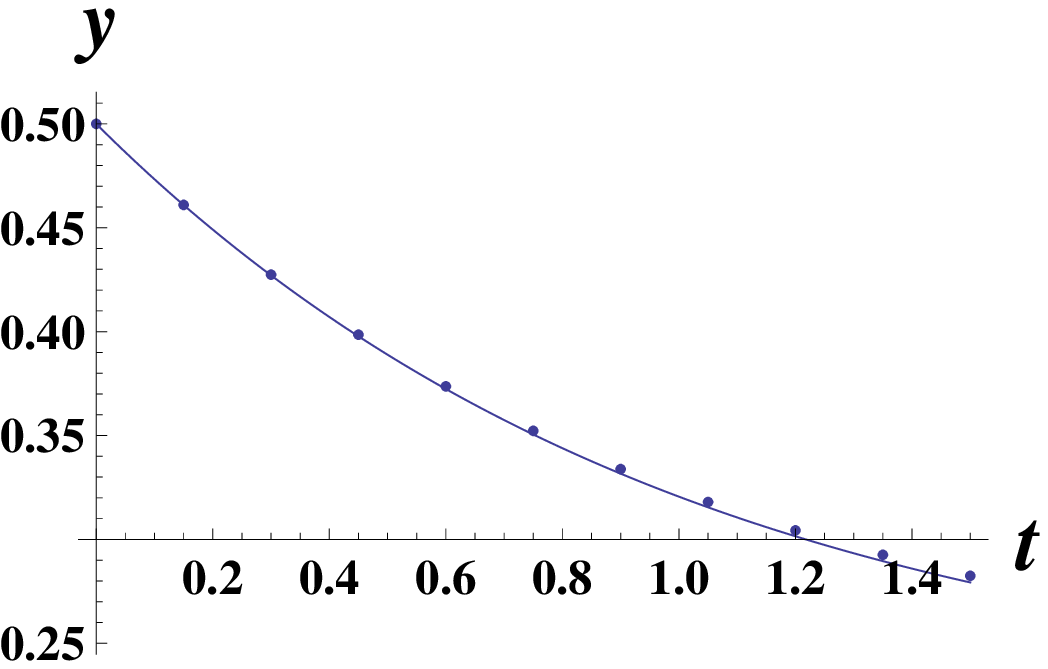}}
\put(2.7,1.0){\small{b.}} \put(2.7,20.5){\small{a.}}
\end{picture}
\vskip 10 mm \caption{The model by Eq.(31), $N=4, \rho =1, h =0.005,
f =0.02$. The direct solution of the system of equations
(smooth line) versus the analytical results (solid circles). $y(t)$
describes the point of maximum,$1/Q(t)$ is the variance of
distribution.}
\end{figure}

\begin{figure}
\centerline{\includegraphics[width=0.95\columnwidth]{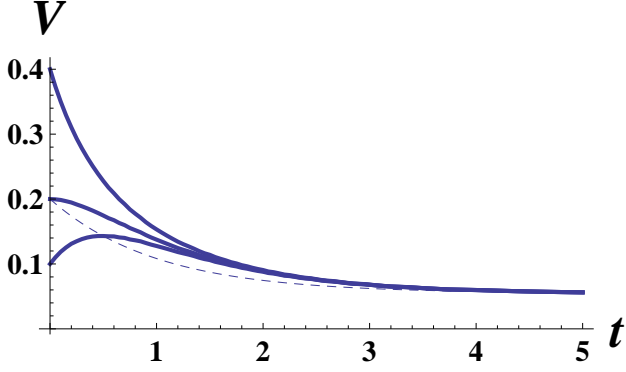}}
\caption{ The   model by Eq.(31), $N=100, \rho =1, h =0.005, f
=0.02$. The numerical result for  $v_2/v_1$  (smooth lines) for
different initial values versus the analytical result (dashed line).
While the numerical results for $y(t)$ and $Q(t)$ actually are
identical with the corresponding theoretical results, regardless of
the initial value of $v_2(0)/v_1(0)$, the ratio $v_2(t)/v_1(t)$
coincides with the theoretical value after some period of time $T$,
where $T\ll N$.} \label{fig1}
\end{figure}

\section{Strip of  D-dimensional systems of equations}

\subsection{The bulk solution of the model} Consider the following system of equations
\begin{eqnarray}
\label{e38} \frac{dP_l(\vec X)}{Ndt}=\sum_{1\le n\le d}\sum_{\vec
m}R_{ln\vec m}(\vec X-\vec m) P_n(\vec X-\vec m,t).
\end{eqnarray}
Here the sum via $\vec m$ is over the vector space of the
stoichiometry vectors of all of reactions.
 Now there are transitions in the same D-dimensional
space with the rate $R_{ll\vec m}$ and transitions between different
D-dimensional spaces with the rates $R_{ln\vec m}, l\ne n$. The
negative $R_{ll0}$ are chosen to ensure the probability conservation
condition.
\begin{eqnarray}
\label{e39}
\sum_{1 \le l \le d} \sum_{\vec m} R_{ln \vec m}(\vec X) = 0
\end{eqnarray}
where at the limit $N\to \infty$
\begin{eqnarray}
\label{e40} R_{ln\vec m}(\vec X-\vec m)=[r_{ln\vec m}(\vec x)+
\frac{\bar r_{ln\vec m}(\vec x)}{N}+O(\frac{1}{N^2})]N^{\bar n}
\end{eqnarray}
where the scaling $\bar n$ is the same for all index  sets
$(l,n,\vec m)$. We can rescale the time, then putting $\bar n=0$.

Assuming an ansatz
\begin{eqnarray}
\label{e41} P_l(\vec X,t)=v_l\exp[Nu(\vec x,t)],
\end{eqnarray}
we have the following system of equations:
\begin{eqnarray}
\label{e42} v_l\frac{\partial u}{\partial t}=\sum_{1\le n\le
d}\sum_{\vec m}r_{ln \vec m} v_ne^{-\sum_k m_ku'_k}.
\end{eqnarray}
For the consistency of the system we again assume that
\begin{eqnarray}
\label{e42} det[M_{ln}(\vec x,\vec u')-q\delta_{ln}]=0,\nonumber\\
M_{ln}(x,u')=\sum_{\vec m}R_{ln\vec m} e^{-\sum_k m_ku'_k},\nonumber\\
q=\frac{\partial u}{\partial t}.
\end{eqnarray}
We assume an ansatz

\begin{eqnarray}
\label{e44}
u(\vec
x,t)=-\frac{1}{2}\sum_{i,j}V_{ij}(x_i-y_i(t))(x_j-y_j(t))
\end{eqnarray}
As before, we consider the decomposition of the determinant via
$H_l,0\le l\le d$. To calculate the dynamic of the maximum, we
neglect the $(-q)^l$ terms for $l$-s higher than 1. Let us
differentiate the expression $H_0-qH_1$ with respect to $x_k$.

\begin{equation}
\label{e44} \sum_{i} \frac{\partial H_0 (\vec x, \vec p)}{\partial
p_i} \frac{\partial p_i}{\partial x_k} - q'_{x_k}H_1=0.
\end{equation}

As $ \frac{\partial p_i}{\partial x_k} = -V_{ki} $ and $q'_{x_k} =
\sum_{i} V_{ki} \frac{dy_i(t)}{dt}$, we have
\begin{eqnarray}
\label{e45}
\sum_{i}V_{ki}(\frac{\partial H_0 (\vec x, \vec p)}{\partial p_i} + \frac{dy_i(t)}{dt}H_1 )=0 \nonumber\\
b_i \equiv \frac{dy_i(t)}{dt} = \frac {-\frac{\partial H_0 (\vec x,
\vec p)}{\partial p_i}|_{\vec p =0}}{H_1(\vec y, 0)}.
\end{eqnarray}

Differentiating the equation $H_0-qH_1+H_2q^2$ with respect to
$x_i,x_j$ , we obtain for the correlation functions:
\begin{eqnarray}
\label{e46}
H_1\frac{d V_{ij}(t)}{dt} + 2H_2\sum_{l} V_{il}b_l \sum_{n} V_{jn}b_{n} \nonumber \\
+ \sum_{l}V_{il}b_l \sum_{n} H'_{1p_n}V_{jn} + \sum_{l}V_{jl}b_l
\sum_{n} H'_{1p_n}V_{in} \nonumber\\+ \sum_{ln} H''_{0p_l p_n}
V_{il} V_{jn}\nonumber\\
- \sum_{l}V_{il}b_l H'_{1x_j} - \sum_{l}V_{jl}b_l H'_{1x_i} \nonumber \\
- \sum_{l}H''_{0x_i p_l} V_{jl} -  \sum_{l}H''_{0x_j p_l} V_{il}
 = 0.
\end{eqnarray}
This is a complete system of equations.

\subsection{The finite N corrections} Let as assume that the finite size corrections are $u_1/N$ for $u$
and $h_l/N$ for $v_l$. Thus, we have:
\begin{eqnarray}
\label{e47} P_l(\vec X,t)=(v_l+\frac{h_l}{N})\exp[Nu(\vec
x,t)+u_1(\vec x,t)]
\end{eqnarray}
The Hamilton-Jacobi equation for the correction terms will be:
\begin{eqnarray}
\label{e48}
v_l \frac{\partial u_1}{\partial t} + h_l \frac{\partial u}{\partial t} = \nonumber\\
\sum_{1 \le n \le d} Q_{ln} v_n + \sum_{n,\vec m}r_{ln\vec m}(\vec x)h_n e^{-\sum_k m_k u'_k} \nonumber\\
Q_{ln} = \sum_{\vec m} (\bar{r}_{ln\vec m}(\vec x) + r_{ln \vec m}(\vec x)*G) \nonumber\\
G = \frac{1}{2}\sum_{i,k=1}^{D}\frac {\partial^2 u}{\partial x_i \partial x_k} m_i m_k - \sum_k m_k \frac{\partial u_1}{\partial x_k}
\end{eqnarray}

From the consistency of $v_l$ system we get the following equation for $u'_1$:
\begin{eqnarray}
\label{e48}
det[Q_{ln}-\frac{\partial u_1}{\partial t} \delta_{l,n}] = 0
\end{eqnarray}
Having the solution for $u'_1$, we define the correction terms
$h_n$.

\subsection{When we need in several $H_l$?}

 In \cite{as11} has been investigated the $D=2,d=2$
model in our classification.

\begin{eqnarray}
\label{e48}\frac{dP_{mn}}{dt}=(A-f(n))P_{mn}+g(n)Q_{mn}\nonumber\\
\frac{dQ_{mn}}{dt}=f(n)P_{mn}+(A+a(E_m^{-1}-1)-g(n))Q_{mn}
\end{eqnarray}
where $E_n^j f(n)=f(n+j)$ and $A=(E_n^{1}-1)n+\gamma
(E_m^{1}-1)m+\gamma b m(E_n^{-1}-1)$.

The authors applied HJE approach to investigate the steady state,
using same Hamiltonian, proportional to  $H_0$ in our notification.
Their Hamiltonian  gives exact equation for the steady state, while
for the dynamic it is only an approximation, and to get correct
dynamics in \cite{as11} has been done some rescaling.

 Our
approach provides the exact dynamics in this case, using $H_1$ for
the dynamics of the maximum, and $H_2$ for the dynamics of the
variance.

For $H_1(p=0)$ we obtain an expression:
\begin{eqnarray}
\label{e49}H_1(p=0,n/N)=f(n)+g(n)
\end{eqnarray}
It is a nontrivial function of $n/N$, therefore $H_0$ alone is not
enough to give a correct dynamics of the maximum.
\section{Conclusion}

In the present article we solved the Chemical Master Equation for
the case of a strip of 1 dimensional chains of equations. Such a
master equation appears in the problem of connected diffusion in
discrete time: diffusion in space direction plus random jumping
between discrete set of states.
 We found a solution for the general
case of the model when the width of the strip is smaller than the
length, and calculated the dynamics of the maximum and the variance
in the large size limit.  We applied our method to get the solution
for the problem of a gene switch model. The numerics confirmers that
the dynamics of the maximum and the variance follows to our
analytical results after very short period of time.

 We also
investigated the case of a d-strip of equation chains, each one
defined in D-dimensional space, and gave a system of ODEs to define
the motion of the maximum point and the dynamics of the variance. In
the appendix we give the ODE systems for the investigation of the
insertion-deletion-base substitution problem in evolution.

We considered the situation when Hamilton Jacobi equation is valid,
assuming that the maximum of distribution for different chains are
at the same point. This was the case in the example considered in
section II-D, satisfying two conditions: a) where the rates $r(x)$
are smooth functions  and b) $d\ll N$. In \cite{sa08} a $D=2$ model
has been considered where condition b) is broken. As a result, the
solution of the system of equations could not be described via the
HJE ansatz of a smooth function $u(x_1,x_2)$.
To investigate the validity of our main assumption Eq.(14), one
needs to undertake further work with more thorough mathematics.
 Because
nowadays the HJE is becoming popular in solving the CME, as well as
for the solution of the diffusion process in case of weak noise,
this issue (the validity of the HJE ansatz) is becoming more
important. Another related fundamental open problem is the limit of
the application of diffusion approximation in master equations of
evolution research \cite{ew04}. While the diffusion approach is a
rather pure approximation in case of the inhomogeneous master
equation \cite{sa07,sa08a}, it is a very closely related technique
to solve the HJE for homogenous master equations (as considered in
the current article), see also \cite{qi11}. Clarifying the limits of
the powerful HJE method could help to clarify the limits of the
diffusion method as well.

{Besides the HJE (WKB in literature) method, quantum mechanical
approach and the moment closing method have been applied to
investigate CME. The advantage of the method: it allows a series
expansion in higher degrees of 1/N, a uniform convergence of our
approximation, which lucks for moment closing approach. Sometimes a
quantum mechanical approach is as effective as the HJE one, but HJE
has wider area of applications.}

 {Let us
briefly discuss some related works. In \cite{sr09} has been solve
the related model, using the generating function method, and has
been derived several impressive results in the large number of
particles limit. This exact method is heavier than HJE approach,
used in current work, and has narrow area of application. In
\cite{mu09} the gene expression has been investigated using spectral
method. Perhaps the method is good for the numerics.
 }

DBS thanks S. Jain and anonymous referee for discussions. This work
was supported by Academia Sinica, National Science Council in Taiwan
with Grant Number NSC 99-2911-I-001-006.

 \renewcommand{\theequation}{A.\arabic{equation}}
\setcounter{equation}{0}
\appendix
\section{Case of two 1-d chains, section II-D}

We consider the model given by Eq. (31). We obtain the following
 analytical solutions for equations Eq.(17) and Eq.(24):
\begin{equation}
y(t) = \frac{k}{\rho} + (y_0-\frac{k}{\rho})e^{-\rho t},
\end{equation}
\begin{equation}
Q(y) = y - y_0\big(\frac{y-k/\rho}{y_0 - k/\rho}\big)^2,
\end{equation}
where $y_0$ is the initial point of the maximum of distribution.

It is also possible to calculate the variance with more precision if
we find higher coefficients for the ansatz of u(x,t). To find T(y)
we use the following equation:
\begin{equation}
\label{ee27} \frac{d^3[q^2H_2-qH_1+H_0]}{dx^3}=0.
\end{equation}
Using the expressions for the higher derivatives of $H_0$ and $H_1$,
we eventually get the following differential equation for T(y) at
$x=y(t)$:
\begin{eqnarray}
\label{ee28}
\frac{dT}{dy}bH_1 = \nonumber\\
-T\big[6Vb^2-3bH'_{1x}+6VbH'_{1p}-3H''_{0xp}+ 3VH''_{0pp}\big] \nonumber\\
- 3\frac{dV}{dy}b(H'_{1x}-VH'_{1p}-2Vb)\nonumber\\
+ 3Vb(V^2H''_{1pp}-2VH''_{1xp}) \nonumber\\
+ 3VH'''_{0xxp} -3V^2H'''_{0xpp} + V^3H'''_{0ppp} {}
\end{eqnarray}
where $V=1/Q,b=dy(t)/dt$.

 \renewcommand{\theequation}{B.\arabic{equation}}
\setcounter{equation}{0}

\section{High dimensional case}
\subsection{The dynamics in D-dimensional space}
For the model given by Eq.(4) we get the HJE for the function
$u(\vec x)$ with following Hamiltonian:
\begin{eqnarray}
\label{ee29} \frac{\partial u}{\partial t}+H(\vec x,\vec
p)=0,\nonumber\\
- H(\vec x,\vec p)=\sum_{\vec n} \big[  R_{\vec n}(N\vec x - \vec n)
e^{-\vec n \vec p} - R_{\vec n} (N\vec x)  \big].
\end{eqnarray}
Let us consider an ansatz
\begin{eqnarray}
\label{ee30} u(\vec
x)=-\frac{1}{2}\sum_{i,j}V_{ij}(x_i-y_i(t))(x_j-y_j(t)).
\end{eqnarray}
Substituting into the above equation, and differentiating by $x_i$,
we get:
\begin{eqnarray}
\label{ee31} \sum_j V_{ij} \frac{dy_j}{dt}=\sum_j V_{ij} H'_{p_j}.
\end{eqnarray}
The last system of equation is consistent at:
\begin{eqnarray}
\label{ee32} \frac{dy_j}{dt}=H'_{p_j}.
\end{eqnarray}

As a result, we have the following system for the $V_{ij}$
\begin{eqnarray}
\label{ee34}
\frac{dV_{ij}}{dt}= \sum_{lm}\frac{\partial^2 H}{\partial p_l \partial p_m} V_{li} V_{mj} - \nonumber\\
-\sum_{l} \frac{\partial^2 H}{\partial p_l \partial x_i} V_{lj}
-\sum_{l} \frac{\partial^2 H}{\partial p_l \partial x_j} V_{li}.
\end{eqnarray}

\subsection{The dynamics for the evolution model with insertions and
deletions}

Consider the following model, describing the insertions and
deletions \cite{sa08a}, $P(N,L)$ is the density of the states having
genome length N and L of $+$ spins, a is the mutation rate, b is the
deletion rate, c is the rate to insert + or - spins:
\begin{equation}
\label{ee35}
\begin{aligned}
&\frac{dp(N,L)}{dt}= \\
& p(N,L) \left( N_0 \,-
\frac{N}{N_0}(a+b)- c \, \frac{N+1}{N_0} \right) \\
&+a \left( \frac{L}{N_0}p(N,L-1)+p(N,L+1)\frac{N-L}{N_0} \right) \\
&+b \left[ p(N+1,L+1)+p(N+1,L) \right] \frac{N+1}{N_0} \\
&+\frac{c}{2} \left( p(N-1,L-1)\frac{L}{N_0}+p(N-1,L)\frac{N-L}{N_0} \right) \\
\end{aligned}
\end{equation}
We should re-write the equation for $P(N,L)=p(N,L) \binom{N}{L}$.
\begin{equation}
\label{ee36}
\begin{aligned}
&\frac{dP(N,L)}{dt}=\\
&P(N,L)\left (N_0 -(a+b)\frac{N}{N_0} -c\frac{N+1}{N_0} \right ) \\
&+a \left ( \frac{N-L+1}{N_0} P(N,L-1) + \frac{L+1}{N_0}P(N,L+1) \right ) \\
&+b \left ( \frac{L+1}{N_0}P(N+1, L+1) + \frac{N-L+1}{N_0}P(N+1,L)  \right ) \\
&+\frac{c}{2} \left ( \frac{N}{N_0}P(N-1,L-1) +\frac{N}{N_0}P(N-1,L).\right ) \\
\end{aligned}
\end{equation}
We have HJE
\begin{eqnarray}
\label{ee37} -H(x_1,x_2,p_1,p_2)=
N_0 -(a+b+c)x_1 + \nonumber \\
+a [ (x_1-x_2)e^{-p_2} + x_2e^{p_2} ] \nonumber\\
+b [ x_2 e^{p_1+p_2} + (x_1-x_2)e^{p_1} ] \nonumber\\
+\frac{c}{2} [ x_1e^{-p_1-p_2} +x_1 e^{-p_1} ].
\end{eqnarray}
We are again interested to find the dynamics of the maximum and the
variance. Let us introduce the following ansatz:
\begin{eqnarray}
\label{ee38}
u(x_1,x_2,t) = -\frac{1}{2}V_{11}(x_1-y_1(t))^2 - \nonumber\\
- \frac{1}{2}V_{22}(x_2-y_2(t))^2 - V_{12}(x_1-y_1(t))(x_2-y_2(t)).
\end{eqnarray}
For the maximum dynamics we get:
\begin{eqnarray}
\label{ee39} \frac{dy_1}{dt}= (c-b)y_1\nonumber\\
\frac{dy_2}{dt}= a(y_1-2y_2)-by_2+\frac{c}{2}y_1.
\end{eqnarray}

The solutions of the latter equation is:
\begin{eqnarray}
\label{ee40}
y_1(t) = y_{10}e^{(c-b)t} \nonumber\\
y_2(t) = y_{20}e^{-(2a+b)t}+\frac{y_{10}}{2}(e^{(c-b)t}-e^{-(2a+b)t}). \nonumber\\
\end{eqnarray}
Thus, we get the following expression:
\begin{eqnarray}
\label{ee41}
\frac{y_2(t)}{y_1(t)} = \frac{y_{20}}{y_{10}}e^{-(2a+c)t} + \frac{1}{2}[1-e^{-(2a+c)t}] \nonumber\\
\lim_{t\rightarrow \infty} \frac{y_2(t)}{y_1(t)} = \frac{1}{2}.
\end{eqnarray}

For the variances we eventually get the following system of
equations:
\begin{eqnarray}
\label{ee42}
-\frac{dV_{11}}{dt} = a[2V_{12}+y_1V_{12}^2] \nonumber\\
+ b[y_2(V_{11}+V_{12})^2 -2V_{11} + (y_1-y_2)V_{11}^2] \nonumber\\
+ \frac{c}{2}[4V_{11}+2V_{12}+y_1(V_{11}+V_{12})^2 + y_1V_{11}^2].
\end{eqnarray}

\begin{eqnarray}
\label{ee43}
-\frac{dV_{12}}{dt} = a[V_{22} -2V_{12}+y_1V_{12}V_{22}] \nonumber\\
+ b[-2V_{12} + y_2(V_{12}+V_{22})(V_{12}+V_{11}) \nonumber \\
+ (y_1-y_2)V_{11}V_{12}] \nonumber\\
+ \frac{c}{2}[2V_{12} + V_{22}+y_1(V_{12}+V_{22})(V_{12}+V_{11}) \nonumber\\
+y_1V_{11}V_{12}].
\end{eqnarray}

\begin{eqnarray}
\label{ee44}
-\frac{dV_{22}}{dt} = a[-4V_{22}+y_1V_{22}^2] \nonumber\\
+ b[-2V_{22}+y_2(V_{12}+V_{22})^2+(y_1-y_2)V_{12}^2] \nonumber\\
+ \frac{c}{2}[y_1(V_{12}+V_{22})^2 + y_1V_{12}^2].
\end{eqnarray}


\begin{thebibliography}{99}

\bibitem{ga}C. W. Gardiner,  Handbook of stochastic methods. Berlin,
Germany: Springer (1985).
\bibitem{ka92} Van Kampen N G,Stochastic processes in Physics and
Chemistry,North-Holland,Elsevier Science, (1992).
\bibitem{qi08}D. A. Beard, H. Qian, Chemical Biophysics: Quantitative
Analysis of Cellular Systems. Cambridge Texts in Biomedical
Engineering. Cambridge Univ. Press, New York (2008).
\bibitem{qi10}P.-Z. Shi and H. Qian, pp. 175-201, in Irreversible Stochastic Processes, Coupled
Diffusions and Systems Biochemistry  J. Feng et al. (eds.),
Frontiers in Computational and Systems Biology, Computational
Biology 15, Springer-Verlag, London (2010)
\bibitem{hu88}G. Hu,
Physica A {\bf 136},607 (1986).
\bibitem{dy94}M. I.
Dykman,  E. Mori and J. Ross,
 J.Chem. Phys. {\bf 100},
5735(1994).
\bibitem{ka04}V. Elgart and A. Kamenev, Phys. Rev. E {\bf 70}, 041106 (2004).
\bibitem{es09}C. Escudero and A. Kamenev, Phys.Rev. E {\bf
79}, 041149 (2009).
\bibitem{as10} M. Assaf and B. Meerson, Phys.Rev. E {\bf 81}, 021116
(2010).

\bibitem{sa11}A. Martirosyan, D.B. Saakian,  Phys. Rev. E {\bf 84},
021122 (2011).
\bibitem{sa07}D. B. Saakian,
Journal of Stat. Physics, {\bf 128},781(2007).
\bibitem{ka07}K. Sato and K. Kaneko, Phys. Rev. E {\bf 75}, 061909 (2007).
\bibitem{sa08a} D. B. Saakian, O. Rozanova, A. Akmetzhanov,
Phys.Rev. E {\bf 78}, 041908(2008).
\bibitem{kr11}T. Antal, P. L.
Krapivsky, arXiv:1105.1157.
\bibitem{iy11}S. Iyer-Biswas, C. Jayaprakash, arXiv:1110.2804.

\bibitem{ev11}J. Venegas-Ortiz, M. R. Evans arXiv:1105.0301.

\bibitem{ra11}A.F. Ramos, G.C.P. Innocentini, J. E. M. Hornos,
Phys.Rev. E {\bf 83},062902 (2011).
\bibitem{la}L.D. Landaw, E. M. Lifshitz, Mechanics, Third Edition: Volume 1 (Course of Theoretical
Physics), Buterworth and Heinemann, Burlington (1976).
\bibitem{ev02} L. C. Evans, {\it Partial Differential Equations},
  AMS (2002).
\bibitem{me98} A. Melikyan, {\it Generalized Characteristics of
First Order PDEs}, Birkh\"{a}user, Boston (1998).
\bibitem{qi11}H. Ge, H.Qian, Analytical Mechanics in Stochastic Dynamics:
Most Probable Path, Large-Deviation Rate Function and
Hamilton-Jacobi Equation, 2011.


\bibitem{hi96}H. Woodcock and P. G.
Higgs, J. Theor. Biol. {\bf 179}, 61 (1996).
\bibitem{ba01}E. Baake and H. Wagner,  Genet. Res. {\bf 78}, 93
  (2001).

  \bibitem{sa08}D. B. Saakian,
Phys. Rev. E,{\bf 78}, 061920 (2008).
  \bibitem{ew04}W. J. Ewens, {\it Mathematical Population Genetics}
  (Springer-Verlag, New York, 2004).

 \bibitem{sr09}S. Iyer-Biswas, F. Hayot, C. Yayaprakash, PRE {\bf
  79},031911(2009).
  \bibitem{mu09}A. Mugler, A.M. Walczak,C.H. wiggins,  PRE
{\bf 80},041921 (2009).
\bibitem{as11}M. Assaf, E.Roberts and Z. Luthey-Schulten, PRL, {\bf
106},248102(2011).

\end{thebibliography}
\end{document}